\documentclass[a4paper,10pt]{article}
\usepackage[dvips]{graphicx}

\begin{document}
\begin{center}
\textbf{Spin Tunneling in Fe$_8$ molecular nanomagnets[Comment on Europhys.
Lett. 47, 722 (1999) by DEL BARCO et al]}\\
\vspace{0.5cm}
G. Bellessa\\
\end{center}
\textit{Laboratoire de Physique des Solides, B\^atiment 510, Université
Paris-Sud, 91405 Orsay, France}\\
\vspace{0.5cm}

Chudnovsky and Tejada have published without asking for my agreement a paper 
in Europhysics Letters (Ref.~\cite{barco}) on an experiment done in my 
research group in Orsay by Del Barco, Vernier and myself. The Fe$_8$ sample 
was furnished by Tejada.

I shall not comment on this behavior, but I want to notice two important
points:

$\bullet$ The first one is about the way the experiment is presented 
in Ref.~\cite{barco} as the observation of mesoscopic quantum
coherence, when it is more simply an EPR experiment that allows to
observe the transitions inside the fundamental doublet of the
Fe$_8$ cluster splitted by the magnetic field perpendicular to the
easy magnetic axis of this one, as described by Korenblit and
Shender~\cite{KS}.

\begin{figure}[!hbp]
\begin{center}
\includegraphics[scale=0.3]{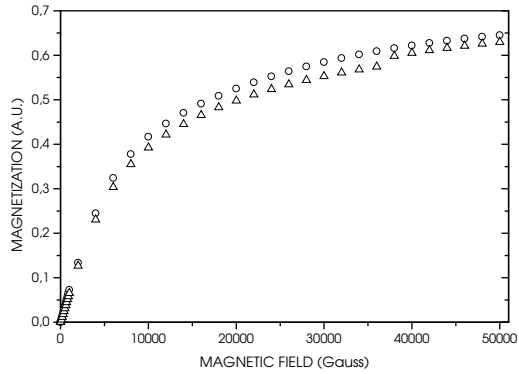}
\end{center}
\caption{\emph{Magnetization of the Fe$_8$ powder at 2~K for the
magnetic field parallel to the orientation axis (the empty
triangles) and perpendicular to the orientation axis (the empty
circles).}}
\end{figure}

$\bullet$ The second point is about the assertion that ``the  
orientation of the powder was done by solidifying an epoxy (Araldit) with 
Fe$_8$ micrometric crystallites buried inside, in a 5.5~T field at 290~K 
during 12 hours''. This assertion is suspicious because the interaction 
energy between magnetic ions inside the Fe$_8$ clusters is much smaller 
than the thermal energy at room temperature. Therefore, I have measured 
the magnetization of \emph{the sample that was used in the experiment of 
Ref.~\cite{barco}} with a SQUID magnetometer at 2~K. The results are 
reported in Fig.~1. It can be seen that there is no significant difference 
between the magnetization curve for the magnetic field parallel to the powder 
orientation axis and the one for the magnetic field perpendicular to the 
powder orientation axis. It is the same in Fig.~1 of Ref.~\cite{barco}. The 
main peaks for the two orientation axes appear different on the high-field 
side but they are similar on the low-field side. So, the difference between 
the two curves is rather due to a change in the base-line.

To conclude, the susceptibility peak reported in Ref.~\cite{barco}
is observed in a non-oriented powder. Only the crystallites that
have their easy axis almost perpendicular to the magnetic field
take part in the susceptibility peak. This explains why the signal
corresponds to a quite small part of the magnetic moments in the
sample.

\end{document}